# A Thermodynamic Model for Active Ion Transport

Xiang Zou, MD

Huashan Hoipital, Fudan University

**Abstract**

Active ion transport is very critical for living cells to maintain and regular internal and external environment. It is known that adenosine 5′-triphosphate (ATP) is a general energy source that applies bond energy for pumps to overcome ion concentration gradient. In this study, we introduce a novel thermodynamic model for active ion transport, which allows the pump to act as a 'Maxwell's demon', and also conforms to the second law of thermodynamics. Transport against the gradient is caused by the thermodynamic fluctuation with information recording in ATP stream. Besides, exhaustive experiments about Na-K ATPase performed in the red cell ghost system are reviewed to verify this model, and all the results can support the solvable theory about the working mechanism of demon pumps. Our findings indicate the possible role of ATP as an information carrier but not energy currency. The high-energy phosphate bonds can improve the efficiency of information recording in relevance to ion transport and may mainly convert to heat.

**Introduction**

It is known that adenosine 5′-triphosphate (ATP) is the universal energy currency in eukaryotic cells. Most of the biological processes in cells rely on the energy released from an ATP to ADP transformation. ATP is mainly synthesized using ADP in mitochondria, and released through ADP/ATP carriers to keep a higher concentration in cytoplasm[1]. As a living cell, it has to keep internal environment different from outer matrix, or a dynamic equilibrium of substance exchange. From the view of thermodynamics, this process is supported by negative entropy intake to resist thermal equilibrium[2].

Ion gradient is the typical thermal unbalance feature of a living cell. The gradient is kept by various ionic pumps, such as Na-K-ATPase, Ca pump and proton pump. The common working mechanism of the pumps is to transport ion from lower concentration side to higher side, consuming ATPs. It is widely believed that high-energy phosphate bonds in ATPs supply the energy maybe result in conformational changes of ionic pumps, achieving ion transport. From a direct speculation, there may be a similar structure in pumps for chemical energy-potential energy transformation, which is not the case. Besides, the ATP consumption among those pumps is just one during one transport, in spite of the ion amount and type differences. All those issues above force us to rethink the accepted knowledge about ionic pumps.

In 1871, J. C. Maxwell raised a gedanken experiment known as 'Maxwell's demon' (J. C. Maxwell, Theory of Heat (Longmans, London, 1871).). In this thought experiment highlighting the statistical nature of the second law of thermodynamics, Maxwell imagined a tiny creature acting as a gatekeeper between two chambers filled with gases at different temperatures. By preferentially allowing fast-moving molecules to pass from the cold to the hot chamber, and slow ones to pass in the other direction, this creature achieves refrigeration without expending energy. The term 'Maxwell's demon' has come to refer not only to the original setting described by Maxwell, but more generally to any situation in which a rectification of microscopic fluctuations produces a decrease of thermodynamic entropy. Taking on this perspective, the ionic pumps are natural demons that can work as Maxwell's idea. In addition, recent progresses in physics have already proved the possibility of Maxwell's demon, not using its intelligence but the stream of information. In this study, we reviewed the exhaustive experiments performed in the red cell ghost system, showing the delicate balance between Na-K pump activation and cell energy currency, and tried to describe the mechanism of ionic pumps from the thermodynamic view, making it easier to understand their complex behavior.

**Materials and Methods**

Experiments related to the red cell ghosts are all reviewed according to the previous descriptions[3,4]. In General, the resealed red cell ghosts were made by procedures involving twice hemolysis. The first hemolysis was carried out to reduce the cellular concentration of endogenous metabolites and adenylate kinase. At the second hemolysis, an adenylate kinase inhibitor was introduced, together with adenine nucleotides and a regenerating system to maintain stable internal ATP and ADP levels. $^{24}$Na was added in resealed process, and other preset concentration of extracellular ion was adjusted[3]. Na efflux rate was calculated by then equation: $\ln(1 - R_s/R_{eq}) = {}_{Na}^{o}K \cdot t$, where $R_s/R_{eq}$ represents the fraction of $^{24}$Na released in time, ${}_{Na}^{o}K$ is the outward rate constant (per hour)[5]. Each point in the plot represents the mean repeated four times in the same condition.

**Results**

**General understanding of the ionic pumps**

Firstly, we introduce an ionic pump that can transfer one ion in one time. The pump consists of two components: a gate as a duplex channel controlled by a certain mechanism; an ATP or ADP binding site, locates in the inner matrix. Outer or inner matrix ion concentrations are defined as $C_h$ and $C_l$, $C_h > C_l$ (Figure 1A). The pump will play the role as Maxwell's demon interacts with the nearest ATP/ADP and with the outer/inner matrix, which will be described in detail as follows.

The pump itself is a two-state system, with states 'f' and 'e' characterized by with and without ion. It can make random transitions between these two states by exchanging ion with the outer matrix (higher ion concentration), as illustrated by the horizontal arrows in Figure 1B. We will refer to these as intrinsic transitions (much higher possibility), to emphasize that they involve the pump but not the ATPs. The corresponding transition rates meet the requirement of detailed balance,

$$\frac{R_{e \to f}}{R_{f \to e}} = e^{-\rho/kC_h}$$

where $\rho$ is the energy difference between two states of the pump, $k$ is Boltzmann's constant.

We assume there are no intrinsic transitions between ATP and ADP. That is, the state of the ATP/ADP can change only via interaction with the pump. At any instant in time, the pump interacts only with the nearest ATP/ADP. During one such interaction interval, the pump and the nearest ATP/ADP can make cooperative transitions: if the ATP combined the pump and the pump is in state 'e', then the pump can simultaneously flip to states f with ATP→ADP, and vice versa (Figure 1B, diagonal arrows). We will use the notation eATP↔fADP to denote these transitions, which are accompanied by an exchange of ion with the inner matrix (lower concentration). The corresponding transition rates again satisfy detailed balance,

$$\frac{R_{\text{eATP} \to \text{fADP}}}{R_{\text{fADP} \to \text{eATP}}} = e^{-\rho/kC_l}$$

For later convenience, we also define,

$$\epsilon = \tanh \frac{\rho(C_h - C_l)}{2kC_hC_l} \qquad (1)$$

whose value $0 < \epsilon < 1$, quantifies the ion concentration difference between outer and inner matrix.

Finally, we assume that the inner matrix contains a mixture of ATPs and ADPs, with proportions $p_t$ and $p_d$, respectively, with no correlations between them. Let

$$\delta \equiv p_t - p_d \qquad (2)$$

denote the proportional excess of ATPs.

We thus have the following dynamics. When ATP/ADP arrives to interact with the pump, it subsequently interacts for a short interval, making the transitions shown in Figure 1B, thus exchanging ion with the matrix. The state of ATP/ADP at the end of the interaction interval is then released to the matrix, and the next ATP/ADP arrives to interact with the pump. The parameter $\epsilon$ also defines the intrinsic and cooperative transition rate difference, and $\delta$ quantifies the incoming ATP/ADP proportion. Under these dynamics, the pump will reach to a periodic steady state, in which its behavior is statistically the same from one interaction interval to the next. In cells, as the amount of pump is considerable, the steady state can be achieved in a very short period.

**How can the pumps transport ion against concentration gradient**

Before proceeding to the solution of these dynamics, we firstly describe how this model can achieve the systematic transfer ion from the lower to the higher concentration side. For this purpose let us assume that each incoming nucleotides are ATPs. At the start of a particular interaction interval, the joint state of the pump and newly arrived nucleotide is either eATP or fATP. The pump and nucleotide then evolve together for an interval, according to the transitions shown in Figure 1B. If the joint state at the end of the interaction interval is eATP or fATP, then it must be the case that every transition eATP→fADP was balanced by an opposite transition. As a result, no ion was absorbed from the lower concentration side. If the final state is eADP or fADP, then we know that there was one ion absorbed from the lower concentration side. Moreover, a record of this process is imprinted in the nucleotide, as every released ADP indicates the absorption of ion from the lower concentration side. Since the pump also exchange ion with the higher concentration side, and since ion cannot be accumulated in the pump, we get a net flux of ion from the lower to the higher concentration side in the long run, related to the amount of ADPs in the released nucleotides. More generally, if the incoming nucleotides contain a mixture of ATPs and ADPs, then an excess of ATPs ($\delta > 0$) induced a statistical bias that favors the ion flow from the lower to higher concentration side, while an excess of ADPs ($\delta < 0$) produces the opposite bias. This bias either competes with or enhances the normal thermodynamic bias due to the concentration difference between the two sides. The pump thus affects the flow of ions between the inner and outer matrix, and modifies the released nucleotides from the pump.

**Ion transport in the frame of second law of thermodynamics**

As described above, we can see that the transport process seems to reduce the entropy without consuming energy. We now investigate this process quantitatively. Once the pump has reached its periodic steady state, let $p'_t$ and $p'_d$ denote the fractions of ATPs and ADPs in the released nucleotides, and let $\delta' = p'_t - p'_d$ denote the excess of released ATPs. Then

$$\Phi \equiv p'_d - p_d = \frac{\delta - \delta'}{2} \tag{3}$$

represents the average production of ADPs per interaction interval in the released nucleotide stream, relative to the incoming binding stream. Since each transition ATP→ADP is accompanied by the absorption of ion from the lower concentration side (Figure 1B), the average transfer of ion from the lower to the higher concentration side, per interaction interval, is given by

$$Q_{l \to h} = \Phi \rho$$

A positive value of $Q_{l \to h}$ indicates that pump transfers ions against a concentration gradient, like the creature imagined by Maxwell.

To quantify the information-processing capability of the demon, let

$$S(\delta) = -\frac{1-\delta}{2}\ln\frac{1-\delta}{2} - \frac{1+\delta}{2}\ln\frac{1+\delta}{2}$$

denote the information content, per nucleotide, of the incoming nucleotide stream, and define $S(\delta')$ by the same equation, for the outgoing bit stream. Then

$$\Delta S \equiv S(\delta') - S(\delta) = S(\delta - 2\Phi) - S(\delta)$$

provides a measure of the extent to which the pump increases the information content of the ATP/ADP pool. A positive value of $\Delta S$ indicates that the pump writes information to the ATP/ADP pool, while a negative value indicates erasure. (More precisely, $\Delta S$ reflects the change in the Shannon information of the marginal probability distribution of released nucleotides).

We see that $\Phi$ determines both $Q_{l \to h}$ and $\Delta S$. As previous description[6], we show that under the dynamics we have described, the pump reaches a periodic steady state, determined by the model parameters $\Lambda \equiv (\delta, \rho, C_h, C_l)$, in which

$$\Phi(\Lambda) = \frac{\delta - \epsilon}{2} \eta(\Lambda), \eta > 0 \tag{4}$$

$$Q_{l \to h} \left( \frac{C_h - C_l}{k C_h C_l} \right) + \Delta S \geq 0 \tag{5}$$

The crucial point is that the sign of $\Phi$ is the same as that of $\delta - \epsilon$. We can think of two effective forces: the bias induced by the incoming nucleotide stream, which favors $\Phi > 0$ when $\delta > 0$, and the concentration gradient, quantified by $\epsilon$, which favors $\Phi < 0$. The direction of ion flow is determined by the difference $\delta - \epsilon$. If $\delta - \epsilon$ equals to zero, that indicates a balance between pumping and inflow forces. In living cells, a stable level of ATP is critical to maintain the ion concentration gradient. On the other hand, a normal ion concentration gradient is just relied on an adequate ATP/ADP ratio.

**Dynamics of Na efflux in the red cell ghost membrane**

In accordance with the theory above, we can see that the transport ability of pumps has little relationship with the energy of nucleotide, but the concentration proportion. For example, in Na-K-ATPase, there are five cations being pumped against the concentration gradient, but only use one ATP, if $\Phi > 0$. Now we calculate the $\Phi$ in details as,

$$\Phi = \frac{\delta - \tanh\frac{\rho(C_h - C_l)}{2kC_hC_l}}{2} \left( 1 - e^{-(1 - \tanh\frac{\rho}{2kC_h}\tanh\frac{\rho}{2kC_l})} \right) \tag{6}$$

In the case of Na-K-ATPase, the existence of potassium can be proportionally equivalent to sodium. We calculated the Na-K exchange rate ($\Phi$) by increasing intracellular ATP or ADP concentration, at various extracellular Na environment (normalized). We found that Na efflux rate is improved by increasing ATP concentration, at various extracellular Na environments (Figure 2A). Besides, Na-K exchange rate can be significantly increased by the removal of extracellular Na, at any ADP concentration (normalized); and the relative ADP concentration can improve the effect induced by extracellular Na removal (Figure 2B). Then we reviewed the effect of varying intracellular ATP on the extent of ouabain-ensitive Na/K exchange in red cell ghosts. We found that Na efflux constant was examined as a function of the ATP concentration, in which the least-squares curve is very similar to our prediction (Figure 2C). Moreover, the percent stimulation of ouabain-sensitive Na efflux caused by removal of external Na was calculated by rate difference divided by the rate in the situation of Na existence. As internal ADP was increased the percent stimulation was raised in step as we estimated (Figure 2C). Figure 2E and 2F are the illustrations about those two effects.

In addition, we found that efflux rate is inhibited by increasing ADP concentration, at any extracellular Na environment (Figure 3A). The similar result can also be found in the situation of various intracellular ATP/ADP environments, by increasing extracellular Na concentration (Figure 3B). Then we reviewed the effect of varying the concentration of intracellular ADP, at constant ATP, on ouabain-sensitive Na/K exchange. The phosphoarginine regenerating system was used to control ADP and ATP levels at preset concentrations. The ATP concentrations are slightly different in each condition. In any one experiment, each value for the ATP concentration in the ghosts was within 15% of the mean for the particular experiment. Then we got the significant negative correlation between the ouabain-sensitive flux and increasing ADP (Figure 3C). Similarly, inhibition of ouabain-sensitive Na efflux by increasing extracellular Na concentration was also reviewed at a controlled ratio of ATP/ADP. The ouabain-sensitive Na efflux rate was determined in media containing various concentrations of Na. Straight lines connect points determined in a single experiment. The phosphoarginine regenerating system was used to control preset ADP and ATP concentrations. We also got the significant negative correlation between the ouabain-sensitive flux by increasing extracellular Na level (Figure 3D). Since the preset Na or ADP/ATP environments might be shifted, each connected lines were presented in different layers, just in accordance with our prediction. Figure 3E and 3F are the illustrations about those two effects.

**Discussion**

In this study, our active ion transport model is in accordance with the context of the second law of thermodynamics. The steady-state change in thermodynamic entropy due to the flow of ions, together with the change in information entropy per interaction interval can be viewed as a modified Clausius inequality, in which the information entropy of a random ATP/ADP stream is explicitly assigned the same thermodynamic status as the physical entropy associated with the transfer of ion. Thus our model provides support for the Landauer's principle, which states that a thermodynamic cost must be paid for the erasure of memory.

Recent study about P-type $Ca^{2+}$ pump revealed that the transport process involves reversible steps such as cation or ATP binding. But the step follows release of ADP and extracellular release of $Ca^{2+}$ is irreversible[7], which is indicated in the demon's behavior in our work. As the thermal fluctuations, active ion transport in our frame is only caused by random ion motion rather than any other forces, including the motion against the concentration gradient, which is quite different from traditional theory. However, the accumulation of fluctuation will finally reach to a concentration balance if there is no other intervention. By relevance to ATPase, the participation of ATP to ADP transformation in this process can be regarded as a 0 to 1 switch. From this point of view, the consumption of ATP during ion transport is just like a writing process into the memory register. ATP and ADP are not the energy molecules but the information carrier molecules. Thus, the energy conversion destination of high-energy phosphate bond in ATP can mainly be heat. This understanding about the active ion transport system will explain why the consumption of ATP is so unanimous in spite of ion type and quantity. On the other hand, the recovery from ADP to ATP in mitochondria is the process as erasure of information. In addition, our model can also predict the reverse process from ADP to ATP by passive transport in Na-K-ATPase[8]. In summary, we have raised a novel and solvable theory about the working mechanism of active transport ion pumps, that mimics the behavior of the 'demon' in Maxwell's thought experiment, which can generate an ion flow against the concentration gradient without external work but information recording. Those systems require only an appropriate proportion of ATPs into which information can be recorded as ADPs.

**Figure Legends**

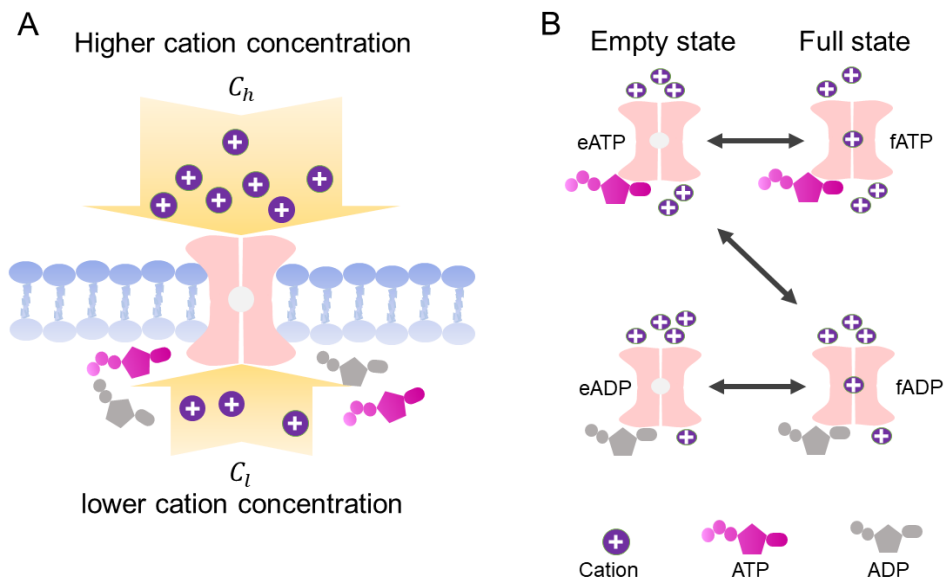

Figure 1. Working mechanism of an active ion transport pump. (A) framework of the pump that can interact with nucleotide, and exchange ion with either lower or higher concentration side. (B) The pump performs intrinsic transitions mediated by the higher concentration side (horizontal arrows). The pump and nearest nucleotide make cooperative transitions eATP→fADP mediated by the lower concentration side (diagonal arrows).

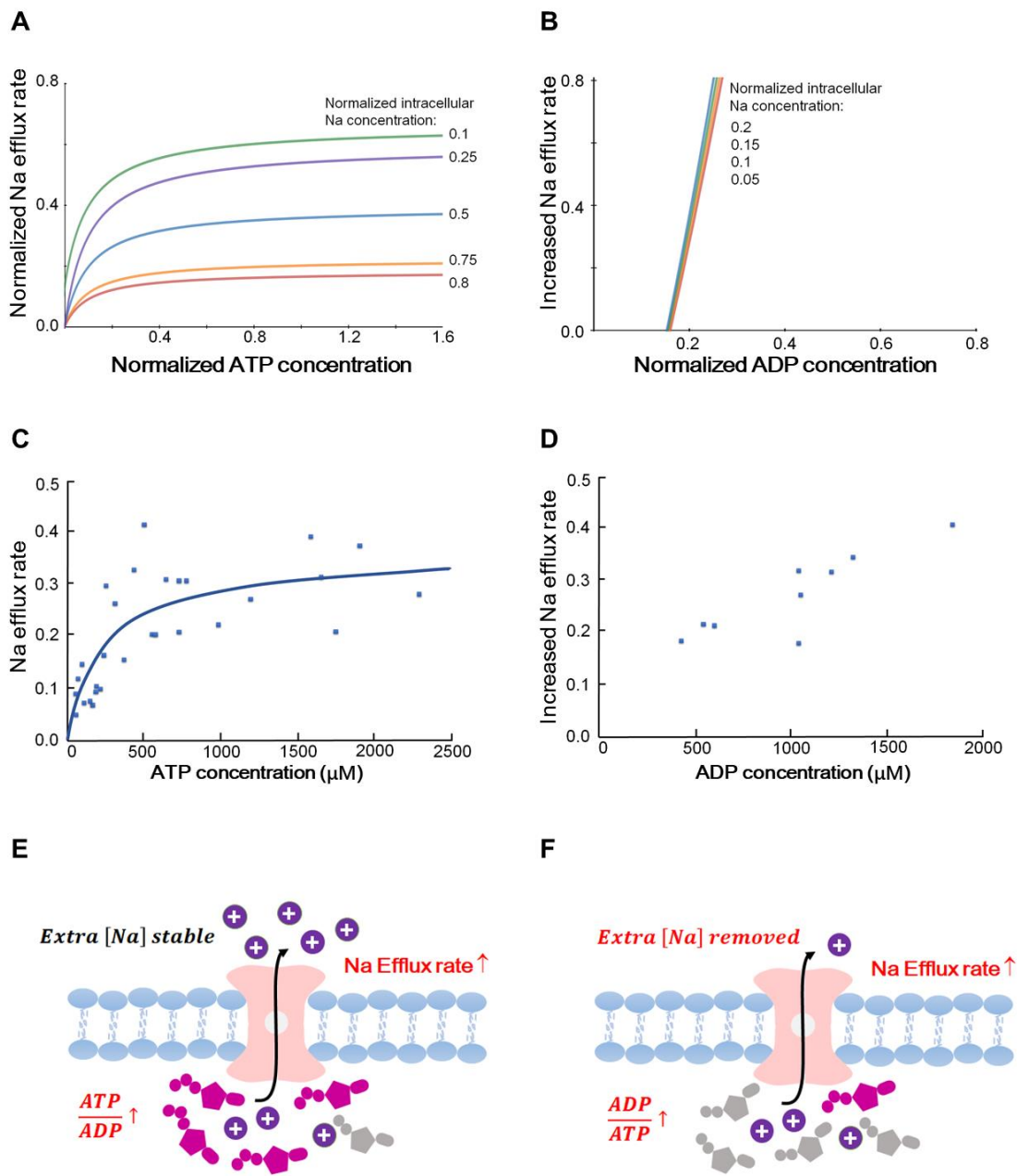

Figure 2. Na-K exchange rate affected by intracellular ATP and removal of extracellular Na. (A) normalized Na efflux rate by increasing ATP concentration, at various normalized preset intracellular Na concentration. Normalized ADP=0.2; extracellular Na=0.8. (B) Increased Na efflux rate after removal of extracellular Na, by increasing ADP concentration, at various normalized preset intracellular Na concentration. Normalized ATP=0.2; extracellular Na=0.8. (C) The effect of various intracellular ATP on the ouabain-sensitive Na-K exchange in resealed ghosts. (D) The stimulation percentage of ouabain-sensitive Na efflux after removal of external Na, by increasing intracellular ADP. (E) - (F), illustrations about those two effects.

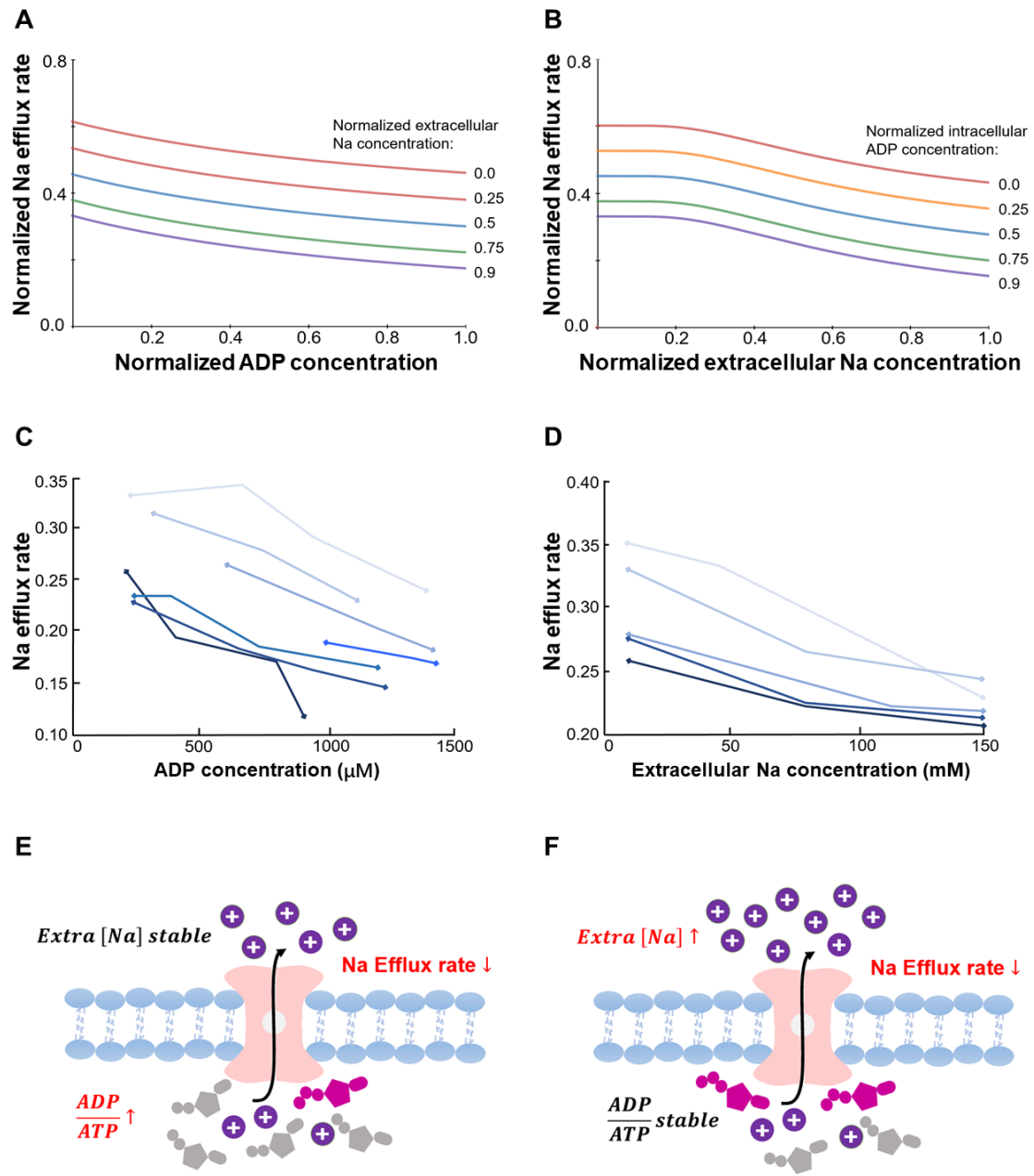

Figure 3. Na-K exchange rate affected by varying intracellular ADP and extracellular Na. (A) normalized Na efflux rate by increasing ADP concentration, at various normalized preset extracellular Na concentration. Normalized ATP=0.8; intracellular Na=0.2. (B) normalized Na efflux rate by increasing extracellular Na concentration, at various normalized preset intracellular ATP concentration. Normalized ATP=0.8; intracellular Na=0.2. (C) The effect of various concentration of intracellular ADP, at constant ATP, on ouabain-sensitive Na-K exchange. (D) Inhibition of ouabain-sensitive Na efflux by increasing extracellular Na at a controlled preset ratio of ATP/ADP. (E) - (F), illustrations about those two effects.